\documentclass[%
 reprint,
 amsmath,amssymb,
 aps,
]{revtex4-2}

\usepackage{graphicx}% Include figure files
\usepackage{dcolumn}% Align table columns on decimal point
\usepackage{bm}% bold math
\begin{document}

\preprint{APS/123-QED}

\title{Constraints on probing quantum coherence to infer gravitational entanglement}


%\author{Onur Hosten}
% \altaffiliation{%
% Institute of Science and Technology Austria, 3400 Klosterneuburg, Austria}%Lines break automatically or can be forced with \\
%  \email{onur.hosten@ist.ac.at}
\author{Onur Hosten}%
\email{onur.hosten@ist.ac.at}
\affiliation{%
 Institute of Science and Technology Austria, 3400 Klosterneuburg, Austria
 }%


\date{\today}% It is always \today, today,

\begin{abstract}

Finding a feasible scheme for testing the quantum mechanical nature of the gravitational interaction has been attracting an increasing level of attention. Gravity mediated entanglement generation so far appears to be the key ingredient for a potential experiment. In a recent proposal [D.  Carney  et  al., Phys. Rev. X Quantum 2, 030330 (2021)] combining an atom interferometer with a low-frequency mechanical oscillator, a coherence revival test is proposed for verifying this entanglement generation. With measurements performed only on the atoms, this protocol bypasses the need for correlation measurements. Here we explore formulations of such a protocol, and specifically find that in the envisioned regime of operation with high thermal excitation, semi-classical models, where there is no concept of entanglement, also give the same experimental signatures. We elucidate in a fully quantum mechanical calculation that entanglement is not the source of the revivals in the relevant parameter regime. We argue that, in its current form, the suggested test is only relevant if the oscillator is nearly in a pure quantum state, and in this regime the effects are too small to be measurable. We further discuss potential open ends. The results highlight the importance and subtleties of explicitly considering how the quantum case differs from the classical expectations when testing for the quantum mechanical nature of a physical system.

\end{abstract}

\maketitle

%\tableofcontents

\section{Introduction}

Experimentally testing the quantum mechanical nature of the gravitational field has long been a subject of interest~\cite{DeWitt2011,Zeh2011,Page1981,Penrose1996}. With recent advances in the quantum mechanical control of mesoscopic objects, the topic has gained renewed interest in the context of optomechanical~\cite{Bose2017,Carney2019,Marletto2017,Kafri2014,Kafri2015,Chevalier2020,Anastopoulos2020,Matsumura2020,Krisnanda2020} as well as atomic systems~\cite{Lindner2005,Howl2021,Haine2021}. Central to the progress is the realization that if two quantum systems can be entangled through the gravitational interaction, the gravitational field ought to be able to transmit quantum information, and thus needs to be described by quantum mechanics~\cite{Bose2017,Marletto2017,Marshman2020,Belenchia2018,Christodoulou2019,Galley2020}. Proposed tests typically require verifying that entanglement is formed between the involved systems by characterizing their correlations. In Ref.~\cite{Carney2021}, a protocol is proposed to bypass this requirement to infer the quantum nature of the interaction from only one of the sub-systems. This would in principle allow one to make the second system very massive while eliminating the need for its quantum control. In particular, the concrete implementation of the protocol  -- referred to as \emph{interactive information sensing} -- aims to utilize an atom interferometer and a macroscopic mechanical oscillator, and to observe the periodic coherence collapse and revivals in the atom interferometer as the signature of dynamic entanglement generation between the two systems. It is argued that the revivals are only possible if the gravitational interaction forms an entangling channel. 

Here we explore formulations of such a protocol both in the semi-classical and in the fully quantum domains, and explicitly illustrate that entanglement is not responsible for the coherence revivals in the configurations required for experimental feasibility -- namely, the high thermal excitation requirement. This emphasizes the importance and subtleties of explicitly considering how the quantum case differs from the classical one for potential experiments aiming at testing the quantum mechanical nature of the gravitational interaction. Further, the results highlight the quantum-to-classical transition where the origin of the loss and gain of quantum coherence changes nature, going from entanglement to classical correlations.

\begin{figure}[b]
\includegraphics{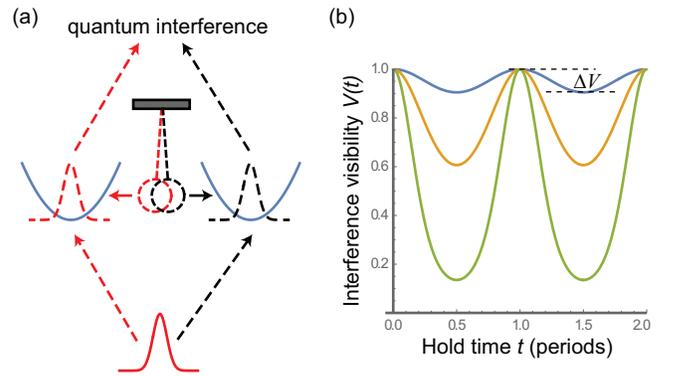}
\caption{\label{fig:epsart} Schematic illustration of the ‘interactive information sensing’ protocol. (a) A quantum superposition of trapped atomic wave packets interact with a mechanical oscillator through gravity. The mechanical oscillator is depicted as a pendulum. (b) Qualitative hold time dependence of atomic interference visibility for increasing (blue, orange, green in order) coupling between the systems. Visibility variations   are exaggerated for illustrative purposes.}
\end{figure}

\section{The protocol}

The protocol is schematically illustrated in Fig.~\ref{fig:epsart}. The wavefuction of an atom is first split into two parts. The resulting wavepackets are trapped (e.g, in the motional ground state of an external potential) and held near the mechanical oscillator. Through the gravitational interaction, the oscillator is pulled to opposite sides depending on the atomic wavepacket location, potentially generating entanglement between the two systems. The expectation is that the systems will repeatedly become entangled and then again separable at the period of the mechanical oscillation. Therefore, depending on the exact instance the atomic wavepackets are released and subsequently interfered, the visibility of the interference fringes formed by the atoms are expected to show periodic collapses and revivals due to the varying level of entanglement with the mechanical oscillator. The measurement of the collapses and revivals form the basis of the proposed scheme. 

As one would expect, the gravitational interaction between the atoms and the mechanical oscillator is quite weak, and thus it is not easy to generate an appreciable amount of entanglement. Concretely, if one assumes a ~10-mg oscillator with a long period of ~100 s, one obtains an interference visibility variation of $\Delta V\sim10^{-24}$ for a mechanical oscillator initially prepared in its ground state. It is clearly not feasible to measure this. However, the authors of Ref.~\cite{Carney2021} notice that, the use of thermal states of the mechanical oscillator significantly improves the variations in the fringe visibility making it potentially feasible to observe it experimentally. In particular for a room temperature oscillator (mean thermal excitation number of $\bar{n}\sim10^{15}$) one obtains $\Delta V\sim10^{-10}$ (and $\Delta V\sim10^{-5}$ for a version of the protocol that incorporates initial entanglement by means of interactions other than gravity). There might be a chance of observing such a visibility variation with an ambitious 100-s atom interferometer with $\sim10^{11}$ atoms with carefully considered geometries, and analyses of noise and systematic effects, as suggested in Ref.~\cite{Carney2021}.

The counter-intuitive aspect of the protocol is that, although the visibility variations improve for initial thermal states with higher temperatures, they fail to improve if the initial oscillator states are taken to be highly excited pure coherent states. It is known that thermal states can be considered as mixtures of coherent states. This suggests that in the thermally excited limit, the visibility collapses and revivals could admit a more mundane classical explanation other than entanglement -- contrary to the theorem provided in Ref.~\cite{Carney2021}. In the remainder of this article, we show that this interpretation is indeed the case. We first show this in the context of a semi-classical calculation, and then we lay out the full quantum mechanical calculation to show that entanglement is not responsible for the coherence revivals in the thermal case. We close by discussing potential open ends. Regarding the possible use of initially entangled systems for enhancing the protocol, we refer the reader to a separate experimental demonstration~\cite{Pal2021} illustrating the interpretational problems that might arise in this context.

\section{Semi-classical calculation}

We take a classical harmonic oscillator with mass $M$ and angular oscillation frequency $\omega$ whose position is given by $x_c(t)$. We place an atom of mass $m$ in an equal superposition of two locations $\pm d$ away from the equilibrium position of the oscillator (Fig.~\ref{fig:epsart}), tightly trapped in the motional ground states of the trapping potentials (corresponding to atomic states that we label $|+\rangle$ and $|-\rangle$). The gravitational potential energies for the two configurations are given by ${E_ \pm } = \frac{-G M m}{| \pm d - {x_c}(t)|} \sim \frac{-G M m}{d} (1 \pm \frac{{x_c}(t)}{d})$, where $G$ is Newton’s constant. Dropping the first term which does not contain any dynamical variables, the Hamiltonian of the atom exposed to the external field generated by the classical oscillator is thus given by
\begin{equation}
\label{eq:semiH}
H =  - \hbar \chi {x_c}(t){\sigma _z}.
\end{equation}

Here $\sigma_z$ is the Pauli operator ${\sigma _z}| \pm \rangle  =  \pm 1| \pm \rangle$, and the coupling strength is defined by $\chi  = \frac{1}{\hbar} \frac{G M m}{d^2}$. The initial state of the atom is given by $|{\psi _0}\rangle  = \frac{1}{\sqrt 2} \left( {|+ \rangle + |-\rangle } \right)$. The oscillator obeys classical equations of motion, and is initially assumed to be in thermal equilibrium with the environment. For the duration of the experiment, the oscillator is assumed to be decoupled well from its environment, so its position is simply given by ${x_c}(t) = {x_0}\cos \omega t + \frac{1}{m \omega} {p_0}\sin \omega t$. In a particular realization of the experiment, the random initial position $x_0$ and momentum $p_0$ of the oscillator are drawn from a classical thermal distribution at temperature $T$. Under the unitary operator $U(t) = \exp [ - \frac{i}{\hbar} \int_0^t {dt'} H(t')] = \exp [i{\sigma _z}{\phi _{{x_0},{p_0}}}(t)]$ generated by the Hamiltonian in Eq.~(\ref{eq:semiH}), for the particular initial conditions $x_0$ and $p_0$, the atomic state evolves as
\begin{equation}
|\psi (t){\rangle _{{x_0},{p_0}}} = \frac{1}{\sqrt2} \left( {{e^{i{\phi _{{x_0},{p_0}}}(t)}}| + \rangle  + {e^{ - i{\phi _{{x_0},{p_0}}}(t)}}| - \rangle } \right).
\end{equation}

Here ${\phi _{{x_0},{p_0}}}(t) = \frac{\chi}{\omega} \left( {{x_0}\sin \omega t + \frac{1}{m \omega} {p_0}\left( {1 - \cos \omega t} \right)} \right)$. The corresponding density operator for this realization of the experiment is ${\rho _{{x_0},{p_0}}}(t) = |\psi (t)\rangle \langle \psi (t){|_{{x_0},{p_0}}}$. Since the initial oscillator values are drawn from a thermal distribution, the overall state of the atom can be described by a mixture expressed with the density operator
\begin{subequations}
\begin{eqnarray}
\rho (t)  &&= \int {d{x_0}d{p_0}} P\left( {{x_0},{p_0}} \right){\rho _{{x_0},{p_0}}}(t)\\
&&\equiv \frac{1}{2} |+\rangle\langle+| + \frac{1}{2} |-\rangle\langle-| + c| + \rangle\langle  - | + c^*| - \rangle\langle  + |, \hspace{30pt} \nonumber\\
&& \hspace{30pt} c = \frac{1}{2} \exp \left[ - 8\left(\frac{\chi}{\omega}\right)^2 \frac{k_B T}{m\omega^2} \sin^2(\omega t/2) \right].\\ \nonumber
\end{eqnarray}
\end{subequations}

Here the thermal distribution probability density is taken as $P({x_0},{p_0}) = \frac{1}{2\pi {\sigma _p}{\sigma _x}} {e^{ - {{p_0^2} \mathord{\left/{\vphantom {{p_0^2} {2\sigma _p^2}}} \right.\kern-\nulldelimiterspace} {2\sigma _p^2}} - {{x_0^2} \mathord{\left/ {\vphantom {{x_0^2} {2\sigma _x^2}}} \right.\kern-\nulldelimiterspace} {2\sigma _x^2}}}}$ with ${\sigma _x} = {{({k_B}T} \mathord{\left/ {\vphantom {{({k_B}T} {m{\omega ^2}}}} \right. \kern-\nulldelimiterspace} {m{\omega ^2}}}{)^{1/2}}$ and ${\sigma _p} = {(m{k_B}T)^{1/2}}$. For a subsequent interference experiment the visibility $V(t)$ of the interference fringes is related to the coherence $c$ as $ V(t) = 2|c|$. In order to compare this result with that of Ref.~\cite{Carney2021}, we define $\lambda  = \frac{\chi}{\omega} {x_{zp}}$, where ${x_{zp}} = {(\frac{\hbar}{2m \omega})^{1/2}}$ is the zero point fluctuations of a quantum harmonic oscillator, and we note that in the highly excited limit ${k_B}T \gg \hbar \omega $ the mean thermal excitation number is given by $\bar n \approx \frac{{k_B}T}{\hbar \omega}$. We thus obtain 
\begin{equation}
\label{eq:visibility}
V(t) = \exp [ - 8{\lambda ^2}2\bar n \sin^2(\omega t/2)].
\end{equation}

The only difference between this equation and Eq. 12 of Ref.~\cite{Carney2021} is the presence of $2\bar n$ instead of $2\bar n +1$. The +1 term is the true quantum contribution as we will see, and is irrelevant in comparison to the intended $\bar n\sim10^{15}$. At half the oscillation period the visibility dips to a minimum and returns to 1 at a full oscillation period, indicating that coherence revivals do take place in a purely semi-classical treatment, without any concept of entanglement. 

Note that semi-classically including the possibility of the atom altering the oscillator motion does not change the discussion. For example in the simplest model, where the gravitational field is sourced by the expectation value of the matter density~\cite{Carney2019}, there isn’t even a force on the oscillator. We now turn to the fully quantum mechanical calculation to see the story from that perspective.

\section{Fully quantum calculation}

Here we include the oscillator in the analysis as a quantum object. With $x$ and $p$ as the position and momentum operators for the oscillator, the Hamiltonian in Eq.~(\ref{eq:semiH}) becomes
\begin{eqnarray}
H &&= \frac{1}{2M}{p^2} + \frac{1}{2} M {\omega ^2}{x^2} - \hbar \chi x{\sigma _z}\\
 &&\equiv \hbar \omega ({a^\dag } - \frac{g}{\omega} {\sigma _z})(a - \frac{g}{\omega}{\sigma _z}).\nonumber
\end{eqnarray}

Above, we defined the creation and annihilation operators through the relation $a = \frac{1}{2{x_{zp}}} (x +i \frac{1}{M \omega} p)$, introduced the scaled coupling constant $g = \chi {x_{zp}}$, and omitted constant energy terms in the conveniently grouped second line. The time evolution of the composite system is then carried out with the unitary operator
\begin{eqnarray}
U(t) &&= \exp [ - i\omega ({a^\dag } - \lambda {\sigma _z})(a - \lambda {\sigma _z})t]\\
&&\equiv {D^\dag }( - \lambda {\sigma _z}){e^{ - i\omega {a^\dag }a t}}D( - \lambda {\sigma _z}).\nonumber
\end{eqnarray}

Here $\lambda  = g/\omega$, and the second expression breaks the evolution operator into a convenient form with the aid of the displacement operator $D(\zeta ) = \exp [\zeta {a^\dag } - {\zeta ^\dag }a]$. Since we are interested in thermal states of the oscillator (which are mixed states), we use the density operator from the outset, and take the joint initial density operator to be a separable one $\rho (0) = {\rho _A}(0) \otimes {\rho _O}(0)$ with
\begin{subequations}
\begin{eqnarray}
{\rho _A}(0) &&= \frac{1}{2}\left( {\mathbb{I} + {\sigma _x}} \right) = \frac{1}{2}\left( {\mathbb{I} + | + \rangle \langle  - | + | - \rangle \langle  + |} \right), \hspace{10pt}
\\
{\rho _O}(0) &&= \int {{d^2}\alpha } \frac{\exp [-|\alpha {|^2}/\bar n]}{\pi \bar n} |\alpha \rangle \langle \alpha |.
\end{eqnarray}
\end{subequations}
The initial atomic state ${\rho _A}(0)$ is a pure state, whereas the initial oscillator state ${\rho _O}(0)$, expressed in the Sudharsan P-representation (in terms of coherent states $|\alpha \rangle$), is a mixed one~\footnote{Note that $Tr[{\rho _O}^2(0)] = \frac{1}{2 \bar n + 1} \to 0$ as $\bar n \to \infty$, indicating a completely mixed state at high temperatures.}. The state evolves as~\footnote{In obtaining this result, we used the following helpful relations about coherent states and displacement operators: $\langle \beta |\alpha \rangle  = {e^{ -|\alpha  - \beta |^2 /2}}{e^{({\beta ^ * }\alpha  - \beta \,{\alpha ^ * })/2}}$, $|\alpha \rangle  = D(\alpha )|0\rangle$, $D(\alpha )D(\beta ) = {e^{(\alpha {\beta ^ * } - {\alpha ^ * }\beta )/2}}D(\alpha  + \beta )$, and ${D^\dag }(\alpha ) = D( - \alpha )$.}
\begin{eqnarray}
\label{eq:density}
\rho (t) =&&\int {{d^2}\alpha } \frac{\exp [-|\alpha {|^2}/\bar n]}{\pi \bar n}\\
&&\times \frac{1}{2} \left( \vphantom{{e^{{\alpha ^ * }{\delta ^ * } - \alpha \delta }}}  {| + \rangle \langle  + | \otimes |{\alpha _ + }\rangle \langle {\alpha _ + }| + | - \rangle \langle  - | \otimes |{\alpha _ - }\rangle \langle {\alpha _ - }|} \right.\nonumber\\
 &&\hspace{40pt} \left. +{{e^{{\alpha ^ * }{\delta ^ * } - \alpha \delta }}| + \rangle \langle  - | \otimes |{\alpha _ + }\rangle \langle {\alpha _ - }| + h.c.} \right).\nonumber
\end{eqnarray}

Here we defined ${\alpha _ \pm } = \alpha {e^{ - i\omega t}} \mp \delta$ with $\delta\equiv\delta(t)  = \lambda ({e^{ - i\omega t}} - 1)$, and $h.c.$ refers to the Hermitian conjugate of the term that precedes it. Since the oscillator goes unobserved, we are interested in the reduced density operator for the atom ${\rho _A}(t) = T{r_O}[\rho (t)]$, obtained by tracing over the oscillator~\footnote{For example, the calculation for $c \equiv \langle  + |T{r_O}[\rho (t)]| - \rangle$ goes as $c = \frac{1}{2} \int {{d^2}\alpha } \, \frac{\exp [-|\alpha {|^2}/\bar n]}{\pi \bar n} {e^{{\alpha ^ * }\,{\delta ^ * } - \alpha \,\delta }}\langle {\alpha _ - }|{\alpha _ + }\rangle $. Given, $\langle {\alpha _ - }|{\alpha _ + }\rangle  = {e^{ - 2|\delta {|^2}}}{e^{{\alpha ^ * }{\delta ^ * } - \alpha \,\delta }}$, this integral results into the expression given in the text.} 
\begin{subequations}
\begin{eqnarray}
{\rho _A}(t) = \frac{1}{2}|+ \rangle &&\langle+| + \frac{1}{2}|- \rangle \langle  - | + c| + \rangle \langle  - | + c^*| - \rangle \langle  + |, \hspace{20pt}
\\
c =&& \frac{1}{2} \exp [-8{\lambda ^2}(2\bar n + 1) \sin^2(\omega t/2)].
\end{eqnarray}
\end{subequations}

This result once again shows the collapses and revivals of the coherence, this time with the quantity $2\bar n +1$ appearing instead if $2\bar n$. To assess the relation of the revivals to entanglement, we note that for small displacements $|\delta | \ll 1$, the coherent states ${|\alpha _ \pm }\rangle$ can be re-expressed exactly as $|{\alpha _ \pm }\rangle  = {e^{-|\delta |^2 /2}}{e^{ \pm ({\alpha ^ * }{\delta ^ * } - \alpha \,\delta)/2}}|\alpha {e^{ - i\omega t}}\rangle \, + {(1 - {e^{ - |\delta {|^2}}})^{1/2}}|{\alpha _{ \bot  \pm }}\rangle$ by utilizing the non-orthogonal nature of coherent states. Here $|{\alpha _{ \bot  \pm }}\rangle$ is some state that is orthogonal to $|\alpha {e^{ - i\omega t}}\rangle$, whose form is not important. Rewriting Eq.~(\ref{eq:density}) using the new expressions for ${|\alpha _ \pm }\rangle$, we obtain the following exact expression for the density operator as a sum of two terms
\begin{equation}
\label{eq:separable}
\rho(t) = {e^{ - |\delta {|^2}}}{\rho ^{(0)}}(t) + (1 - {e^{ - |\delta {|^2}}}){\rho ^{(1)}}(t). 
\end{equation}

Here ${\rho ^{(0)}}$ and ${\rho ^{(1)}}$ both have unit traces. While the exact form of ${\rho ^{(1)}}$ is not important, ${\rho ^{(0)}}$ is given by
\begin{subequations}
\begin{eqnarray}
{\rho ^{(0)}}(t) =&& \int {{d^2}\alpha} \frac{\exp [-|\alpha {|^2}/\bar n]}{\pi \bar n}|{\psi _0}(t)\rangle \langle {\psi _0}(t)| \otimes |\alpha \rangle \langle \alpha|, \hspace{20pt}
\\
|\psi_0(t)\rangle &&= \frac{1}{\sqrt 2} \left( {{e^{({\alpha ^ * }{\delta ^ * } - \alpha \delta )}}| + \rangle + {e^{ - ({\alpha ^ * }{\delta ^ * } - \alpha \delta )}}| - \rangle } \right).
\end{eqnarray}
\end{subequations}

In the case with high thermal excitation, the contribution of ${\rho ^{(1)}}$ to the overall density operator ${\rho}(t)$ is minute (since $\delta  \ll 1$ and ${e^{ - |\delta {|^2}}}\sim1$ in our example), and can safely be ignored. Therefore, to an excellent approximation $\rho (t) \equiv {\rho ^{(0)}}(t)$. In fact $\rho ^{(0)}$ is in the canonical form of a separable mixed state – a sum (or integral) over direct products of the density operators of the subsystems – and hence contains no entanglement. If we calculate the reduced density operator $\rho _A^{(0)}(t) = T{r_O}[{\rho ^{(0)}}(t)]$ for the atoms for this separable state, and in particular the coherence $c^{(0)}=\langle+|\rho _A^{(0)}(t)|-\rangle$, we recover exactly the semi-classical result already obtained in Eq.~(\ref{eq:visibility})
\begin{equation}
{V^{(0)}}(t) = 2|c^{(0)}|= \exp [ - 8{\lambda ^2} 2\bar n \sin^2(\omega t/2)].
\end{equation}

This result shows that for the fully quantum model in the thermally excited limit, the visibility collapses and revivals are not related to the generation of entanglement between the two systems as they are displayed for a fully separable quantum state. An interpretation involving entanglement requires the oscillator to be in a nearly pure state, e.g. nearly at zero temperature ($\bar n\sim0$), or in a coherent state, so that the contribution to visibility variations are verifiably originating from the non-separable aspect of the state. In this pure state limit, although $\rho (t)$ in Eq.~(\ref{eq:separable}) is still dominated by the $\rho^{0} (t)$ contribution, the leading contribution to visibility variations originates from the non-separable $\rho^{1} (t)$ term, relating to entanglement. Thus, only in this limit can visibility variations be attributed to dynamical generation of entanglement.

\section{Discussion}

In summary, we showed that in the high temperature limit of the sensing protocol under discussion, the experimentally observable signatures are identical to those predicted by a semi-classical model. As a consequence, the protocol in this limit is not utilizable in its current form for testing the quantum nature of the gravitational interaction. We further elucidated in the fully quantum mechanical model that indeed the non-monotonic visibility variations in the high temperature limit are well described by a completely separable -- unentangled --  quantum state, showing the explicit identification of the quantum and the semi-classical parts of the problem. Note that although our analysis does not indicate any contradiction with utilizing the protocol for nearly pure initial states of the oscillator, experimental feasibility of observing visibility variations in the coupled atom-oscillator system requires high initial oscillator temperature (with $\bar n\sim{10^{15}}$), rendering the oscillator astronomically far from the pure state limit. The results also highlight the quantum-to-classical transition where the origin of the non-monotonic coherence variations change nature, transitioning from entanglement-driven near the ground state of the oscillator to classical-correlation-driven for highly excited thermal states.

In Ref.~\cite{Carney2021}, a theorem is proven for the necessity of entanglement generation for observing coherence revivals. The regime of validity of this theorem does not cover the semi-classical limit discussed in this work. To hint at the contradiction, we note that the assumptions in the proof for restricting the dynamical evolution into the Lindblad form also restrict the variations of the coherence to a simple exponential decay. This limited form can capture neither the initial quadratic coherence decay nor the coherence revivals in Eq.~(\ref{eq:visibility}) for the perfectly valid semi-classical model described here, which does not involve the concept of entanglement. This highlights that a proof limited to the Lindblad form for dynamical evolution easily misses relevant exceptions.

Subtleties of attributing observable signals in optomechanical systems to classical or quantum origins has been a recurring topic in the analysis of such systems~\cite{Armata2016}. An archetype of a quantum mechanical interaction between two objects and the loss of coherence due to this interaction was first applied to an optomechanical system in the context of the interaction of a movable cavity mirror with a single photon~\cite{Marshall2003}. The protocol discussed here bears stark similarities. A reanalysis of the problem was presented a decade later in the context of a specific hybrid quantum-classical theory~\cite{Lampo2014}, and the findings there resonate with those found here: hybrid quantum-classical theory replicates all system behaviour except when the mirror is nearly in its ground state.

It is not clear if additional assumptions or measurements can rectify the usefulness of the protocol in the high temperature limit. For example, although the full quantum model illustrates the lack of entanglement in the high temperature limit, one could argue that the quantum correlations are still there, and if one hypothetically measured the environment, experimental observability of the entanglement between the oscillator and the atom could be recovered. Nevertheless, knowledge of the outcome of this measurement would project the thermal state of the oscillator into a coherent state, again reducing the conditional visibility variations, taking us back to the experimentally unfeasible regime. On the other hand, for testing the quantum nature of the channel formed by the gravitational interaction, the goal is not necessarily to test the entanglement between the two systems, but to observe any hint that the channel itself has the capacity to entangle, irrespective of formation of any measurable entanglement between the interacting parties. For the tests proposed in Ref.~\cite{Carney2021}, one technically assumes that both the mechanical oscillator and the atom are accepted to behave quantum mechanically, and then one tests for the behavior of the field. Strictly speaking, the semi-classical model for this would require setting up a model where the quantum-oscillator and the atom are both coupled to a third dynamical object - the classical filed. Unlike the semi-classical model elaborated above, this task requires choosing a concrete theory for how to couple a quantum system to a classical one. A consistent example would be the hybrid-classical theory used in~\cite{Lampo2014}. Nevertheless we do not expect the conclusions to change for the high temperature limit.

\emph{Note added} -- The complementary work in Ref.~\cite{Ma2021} recently became available after the initial posting of the this manuscript. Ref.~\cite{Ma2021} discusses semi-classical models and non-classicality as it applies to the protocol under discussion, and provides a more detailed discussion on the restrictiveness of the central theorem of Ref.~\cite{Carney2021}.

\begin{acknowledgments}
OH is supported by IST Austria. The author thanks Jess Riedel for discussions.
\end{acknowledgments}

%\nocite{*}  % show all references including the uncited ones

\bibliography{manuscript}% Produces the bibliography via BibTeX.

\end{document}